\documentclass{iopjournal}
\usepackage[latin9]{inputenc}
\usepackage{todonotes}
\usepackage{amsmath}
\usepackage{amssymb}
\usepackage{graphicx}
\usepackage{esint}
\newif\ifrevision
\revisionfalse    

\ifrevision
  
\else
  
\fi

\makeatletter
\renewcommand\maketitle{}

\let\affilold\affil
\renewcommand{\affil}[1]{\affilold{#1\\}}

\let\articletypeold\articletype
\renewcommand{\articletype}[1]{\articletypeold{#1\\}}

\let\authorold\author
\renewcommand{\author}[1]{\authorold{#1\\}}

\usepackage{color}
\usepackage{dcolumn}
\usepackage{bm}

\makeatother

\begin{document}
\articletype{Paper}
\title{Critical photoinduced reflectivity relaxation dynamics in single-layer Bi-based cuprates near the pseudogap end point}
\author{ T. Shimizu$^{1}$, R. Tobise$^{1}$, T. Kurosawa$^{1}$, S. Tsuchiya$^{1}$,
M. Oda$^{2}$, Y. Toda$^{1,*}$, V. V. Kabanov$^{3}$, D. Mihailovic$^{3}$,
and T. Mertelj$^{3}$ }
\affil{$^{1}$Department of Applied Physics, Hokkaido University, Sapporo
060-8628, Japan}
\affil{$^{2}$Department of Physics, Hokkaido University, Sapporo 060-0810,
Japan}
\affil{$^{3}$Complex Matter Dept., Jozef Stefan Institute, Jamova 39, Ljubljana,
SI-1000, Slovenia}
\affil{$^{*}$Author to whom any correspondence should be addressed.}
\email{toda@eng.hokudai.ac.jp}

\begin{abstract}
A comprehensive study of photoinduced transient reflectivity dynamics
in heavily overdoped single-layer cuprate (Bi,Pb)$_{2}$Sr$_{2}$CuO$_{6+\delta}$
(Pb-Bi-2201) across the end points of the pseudogap and superconducting
phases was conducted using optical ultrafast time-resolved pump-probe
spectroscopy. In the Pb-Bi-2201 near the proposed pseudogap end-point
doping, the transient reflectivity dynamics above $T_{{\rm c}}$ resemble
the pseudogap response observed in the optimally doped La-Bi-2201.
With decreasing temperature, however, the relaxation time exhibits
a power-law divergence, $\tau\sim10\hbar/k_{\mathrm{B}}T$, consistent with
quantum critical behavior near the proposed pseudogap end-point doping. A similar
power-law increase in relaxation time is also observed at a slightly more overdoped composition, though it is less pronounced. 
\end{abstract}
\maketitle

\section{Introduction}

The pseudogap (PG) phase in the cuprate high-$T_{{\rm c}}$ superconductors
exhibits anomalous metallic properties, distinct from those of conventional
Fermi-liquid metals, and has long been investigated as a major factor
in elucidating the mechanism of high-$T_{{\rm c}}$ superconductivity
\cite{timusk1999,keimer2015,sakai2023}. The phase is characterized
by the presence of an incomplete gap structure above $T_{{\rm c}}$
on an energy scale $\Delta_{{\rm PG}}$ reaching up to $\sim80$~meV
\cite{tallon2020locating} in the underdoped regime. With increasing
doping the pseudogap energy scale is monotonically suppressed and
vanishes at the PG end-point doping (PGED) lying in the overdoped
superconducting (SC) phase diagram region.

In recent years, high-quality single crystals of single-layer cuprate
superconductors, such as Pb-doped Bi$_{2}$Sr$_{2}$CuO$_{6+\delta}$
(Pb-Bi-2201) \cite{lin2010,berben2022}, La$_{2-x}$Sr$_{x}$CuO$_{4}$
(La-214) \cite{yoshida2006systematic}, and Tl$_{2}$Bi$_{2}$CuO$_{6+\delta}$
(Tl-2201) \cite{putzke2021}, have become available even in the heavily
overdoped regime, where both the SC and the PG phases vanish. This
advancement has facilitated the investigation of the intrinsic electronic
properties of this regime, offering new insights into the fundamental
nature of high-$T_{{\rm c}}$ superconductivity. Taking the overdoped
Bi-2201 as an example, scanning tunneling microscopy (STM) experiments
have revealed that charge ordering and electronic nematicity persist
in the overdoped regime as a glassy, short-range charge order \cite{Li2018,Li2021}.
Angle-resolved photoemission spectroscopy (ARPES) measurements have
identified a Lifshitz transition in the Fermi surface topology and
suggested its potential connection to the disappearance of superconductivity
\cite{kondo2004holeconcentration,Fujita2014Science,Ding2019,Valla2021}.
Moreover, high-resolution ARPES has provided compelling evidence of
strong electron correlations in the strange-metal state, indicating
the presence of many-body interactions beyond electron-phonon coupling
\cite{Miyai2025}. The presence of ferromagnetic fluctuations has
been suggested by several transport properties and muon spin relaxation
($\mu$SR) experiments~\cite{Kurashima2018,Raffy2022}, consistent
with theoretical predictions \cite{Kopp2007}. Furthermore, transport
experiments have indicated strange metal behavior \cite{putzke2021,girod2021normalstate,Lizaire2021,michon2019thermodynamic,ayres2021incoherent}
in the region of the PGED. While in various overdoped cuprates some
evidence \cite{girod2021normalstate,Lizaire2021,michon2019thermodynamic,tallon1999critical,krusin-elbaum2010interlayer}
points towards quantum criticality at the PGED, the transport properties
and carrier concentration, as well as the SC properties, evolve rather
smoothly across the PGED, suggesting absence of a quantum critical
point (QCP) \cite{putzke2021,ayres2021incoherent,hussey2013generic,hussey2018atale,barivsic2019evidence}.

In this study, we investigate the ultrafast photoinduced nonequilibrium
relaxation dynamics in strongly overdoped Pb-Bi-2201, covering the
overdoped end of the SC dome including the PGED and the putative QCP.
While the ultrafast dynamics has been thoroughly studied in the finite-PG region of the phase diagram \cite{demsar1999superconducting,kusar2005asystematic,toda2011quasiparticle,coslovich2013competition,giannetti2016ultrafast,akiba2024},
the relaxation dynamics data in the strongly overdoped region are
still lacking.

The normal-state \footnote{By the normal state, we refer to the non-superconducting state in
general, including states both below and above the PG crossover temperature.} transient reflectivity in the finite-PG hole-doping region is characterized
by a relaxation component that shows a characteristic $T$ dependence,
set by the PG energy scale $\Delta_{\mathrm{PG}}$ \cite{kabanov1999},
and relaxes on a sub-picosecond to picosecond time scale \cite{demsar1999superconducting,kusar2005asystematic,toda2011quasiparticle,coslovich2013competition,giannetti2016ultrafast,akiba2024,schneider2003carrier}.
While we find that such behavior is reproduced in the optimally doped
La-Bi-2201, the normal-state dynamics in overdoped Pb-Bi-2201 shows
quite different behavior with a divergent relaxation-time slowing
down with decreasing $T$, $\tau\propto T^{-z}$, near the proposed PGED, which is consistent with
a quantum critical scenario in this doping region. As a similar normal-state
transient-reflectivity relaxation slowdown has been observed in electron-doped
cuprates \cite{liu1993ultrafast,long2006femtosecond,cao2008quasiparticle,hinton2013timeresolved,vishik2017ultrafast},
the present data suggest a common underlying physics in heavily hole-
and electron-doped cuprates.


\section{Experimental}

High-quality single-crystal Bi-2201 samples were prepared using the
floating zone method. An optimally doped reference sample (La-Bi-2201;
OPD34) was obtained by partially substituting Sr with La. Very overdoped
(VOD) samples were prepared through partial substitution of Bi with
Pb, followed by annealing under various conditions. The Pb-Bi-2201
sample with $T_{{\rm c}}=10$~K is labeled as VOD10, the one with
$T_{{\rm c}}=7$~K as VOD7, and the non-superconducting Pb-Bi-2201
as VOD0. In this paper, we use ``OD'' solely as a general descriptor
for the overdoped regime in cuprates, while ``VOD'' (very overdoped)
is used as a label for the specific Pb-Bi-2201 samples studied here
(VOD10, VOD7, and VOD0).

\begin{figure}[htbp]
\centering \includegraphics[width=0.6\textwidth]{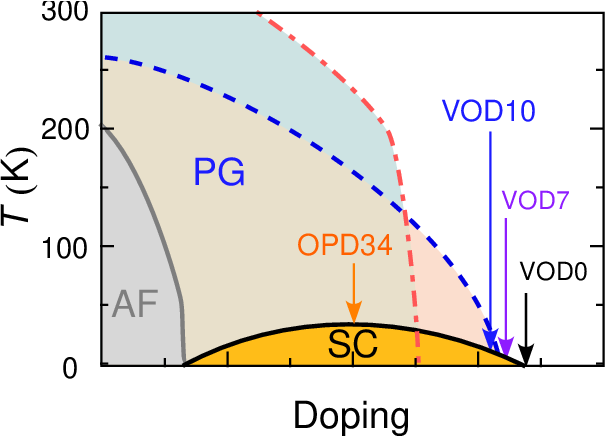} \caption{Temperature-doping phase diagram of Bi-2201. Superconducting (SC)
phase transition temperatures $T_{{\rm c}}^{{\rm m}}$ are 34 K for
OPD34, 10 K for the sample annealed in an Ar atmosphere (VOD10), and
7 K for the as-grown sample (VOD7), while the sample annealed in an
O$_{2}$ atmosphere (VOD0) shows no SC transition down to $T=2.5$~K.
Two suggested pseudogap crossover lines are indicated by the dashed
and dash-dotted lines \cite{berben2022}. }
\label{fig_pd} 
\end{figure}

The correspondence between the samples and the phase diagram is summarized
in Fig.~\ref{fig_pd}, where two suggested pseudogap (PG) crossover
lines are shown (the dashed line and the dash-dotted line), resulting
from different experimental techniques \cite{berben2022}. 
If one adopts the dashed PG crossover line as a working phase diagram, the VOD samples correspond approximately to compositions near the proposed PGED (VOD10), beyond the proposed PGED (VOD7), and beyond the superconducting end-point doping (VOD0), respectively.

Ultrafast optical time-resolved spectroscopy was performed using a
cavity-dumped Ti:Al$_{2}$O$_{3}$ laser (pulse width 120 fs, repetition
rate 270 kHz, $\lambda_{{\rm pr}}=800$ nm) for the probe pulse and
its second harmonic ($\lambda_{{\rm P}}=400$ nm) for the pump pulse.
The sample was mounted on a copper holder within a helium-flow cryostat,
and the coaxially aligned pump and probe beams were focused onto the
sample using a lens. The beam diameter was set to 15~$\mu$m. The
optical permittivity changes induced by the pump pulse carrier excitation
were detected as the probe pulse reflectivity changes with a time
delay $t_{{\rm Ppr}}$, which was controlled by the optical path delay.
To enhance the detection sensitivity, the weak reflectivity changes
of the probe were amplified using a lock-in detection technique, where
the pump pulse intensity was modulated with an optical chopper.


\section{Results}

\begin{figure*}[t]
\centering \includegraphics[width=0.9\textwidth]{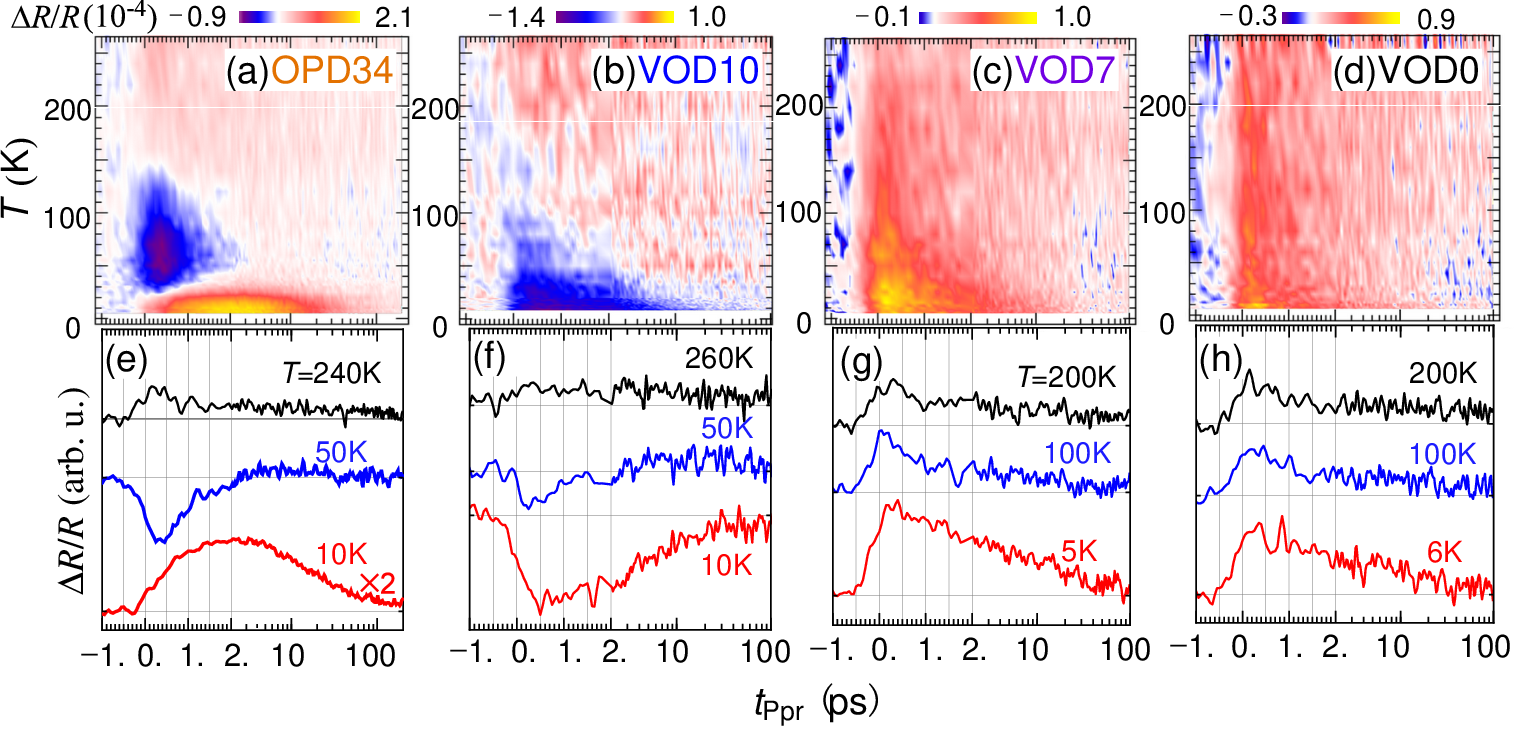} \caption{Temperature dependences of the transient reflectivity ((a--d): color
density plots of $\Delta R/R$, (e--h): $\Delta R/R$ at selected
$T$) for (a, e) OPD34 with the pump fluence of ${\mathcal{F}}_{{\rm P}}=15~\mu{\rm J/cm}^{2}$,
(b, f) VOD10 with ${\mathcal{F}}_{{\rm P}}=2.3~\mu{\rm J/cm}^{2}$,
(c, g) VOD7 with ${\mathcal{F}}_{{\rm P}}=8.5~\mu{\rm J/cm}^{2}$,
(d, h) VOD0 with ${\mathcal{F}}_{{\rm P}}=9.0~\mu{\rm J/cm}^{2}$.
In the color density plots, red and blue colors correspond to positive
and negative $\Delta R/R$, respectively, and white corresponds to
the zero level. All individual $\Delta R/R$ traces underlying the
color density plots (Fig.~S2 in Sec.~B), as well as SC-dominated
data for OPD34 at ${\mathcal{F}}_{{\rm P}}=2.0~\mu{\rm J/cm}^{2}$
(Sec.~D), are provided in the Supplemental Material~\cite{SM}. }
\label{fig_T} 
\end{figure*}

\begin{figure}[t]
\centering \includegraphics[width=0.7\textwidth]{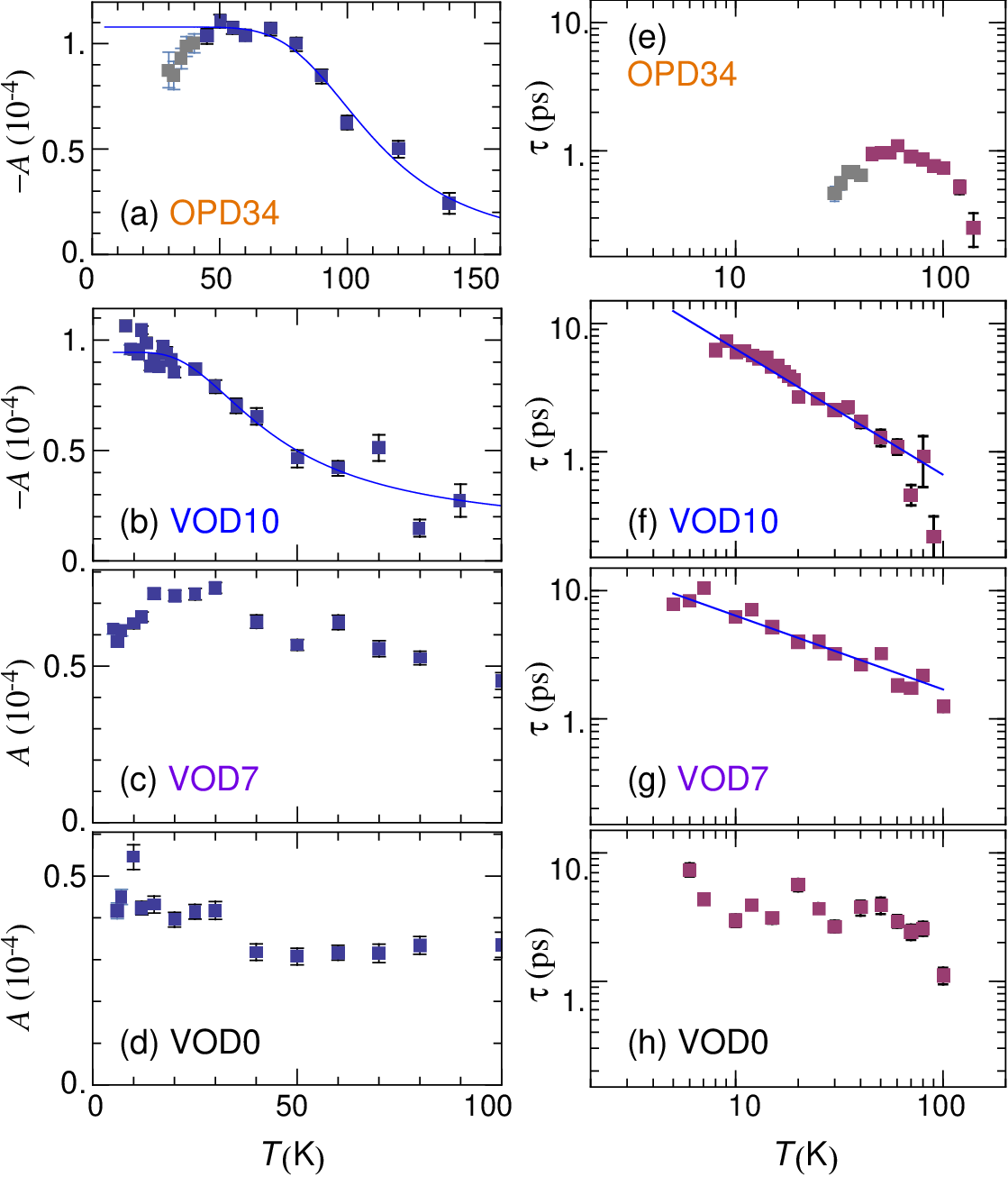} \caption{ Temperature dependence of the amplitude ((a)--(d)) and decay time
((e)--(h)) of the transient reflectivity in (a, e) OPD34, (b, f)
VOD10, (c, g) VOD7, and (d, h) VOD0. For OPD34, (a, e), only the results
above $T_{\mathrm{c}}$ are shown. The solid lines in (a) and (b)
represent fitting curves based on a temperature-independent gap model
\cite{kabanov1999}. Solid lines in (f) and (g) indicate fits of the
relaxation time to a power-law function, $\tau\propto T^{-z}$. }
\label{fig_Tanl} 
\end{figure}

The $T$ dependence of the transient reflectivity in all samples is
summarized in Fig. \ref{fig_T} as color density plots (upper) and
the transient reflectivity traces at representative temperatures (lower).
Figure \ref{fig_T}(a) and (e) show the results for the optimally
doped sample (OPD34), while (b)--(d) and (f)--(h) correspond to
the overdoped samples (VOD$x$). The excitation fluence for each sample
is adjusted according to the transient-response saturation threshold,
into the fluence-linear amplitude region, as detailed in Sec.~C of
the Supplemental Material~\cite{SM}. Under these conditions, local
heating effects in the linear regime are negligible, remaining within
$\sim$1~K (see Fig.~S6(c) of the Supplemental Material). The OPD34
sample exhibits a typical transient reflectivity response that reflects
the characteristics of the phase diagram (Fig. \ref{fig_pd}), as
shown in Fig. \ref{fig_T}(e), where the carrier dynamics in the metallic
(top), pseudogap (middle), and superconducting (bottom) states can
be identified~\cite{akiba2024,liu2008}. A notable feature of the
Bi-based cuprate superconductors is the negative sign \cite{coslovich2013competition,akiba2024,liu2008}
of the near-infrared transient reflectivity relaxation component associated
with the PG. Similar behavior is observed in the VOD10 sample while
the VOD7 and VOD0 samples exhibit the transient reflectivity with
a positive sign over the entire temperature range. In all VOD$x$
samples the low-$T$ non-SC-state relaxation dynamics appears significantly
slower in comparison to the OPD34 sample. Moreover, there is no notable
transient-reflectivity change when entering the SC state in the VOD7
and VOD10 samples.

The $T$-dependent transients are analyzed by fitting them with a single exponential decay function, ${\displaystyle \Delta R/R=A\exp{\left(-t/\tau\right)}+C}$, which provides an excellent description of the measured transients over the entire temperature range (see Fig.~S3 in the Supplemental Material~\cite{SM}).
The amplitude ($A$) and relaxation time ($\tau$) are shown as functions of $T$ in Fig.~\ref{fig_Tanl}. We focus on the data below $T=150$~K for the OPD34 sample in Fig.~\ref{fig_Tanl}(a, e) and below $T=100$~K for the VOD$x$ samples in Fig.~\ref{fig_Tanl} (b)--(d) and (f)--(h).

For OPD34, the PG carrier dynamics dominate the high-temperature side
above $T_{{\rm c}}$. In Fig.~\ref{fig_Tanl}(a), the amplitude of
OPD34 increases gradually with decreasing $T$ and saturates below
$T=70$~K. The decrease observed below $T=45$~K can be attributed
to the contributions from the SC response and its fluctuating component,
which exhibits an opposite sign compared to the PG response. For temperatures
above $T\sim45$~K, the $T$ dependence of the OPD34 transient-reflectivity
amplitude is well described by a $T$-independent gap bottleneck model
\cite{kabanov1999}, previously used to describe the PG component
in various cuprate superconductors. The solid line in Fig.~\ref{fig_Tanl}(a)
represents the best-fit result, yielding the PG energy scale, $\Delta_{{\rm PG}}=52\pm6$
meV. It should be noted that, during the fitting procedure, the positive
metallic response, observed at high-$T$, is assumed to remain constant
throughout the entire $T$ range under consideration.

In Fig.~\ref{fig_Tanl} (b), the $T$ dependence of the amplitude
in the VOD10 sample is qualitatively similar to that in the OPD34
sample, showing a gradual increase followed by a plateau as the temperature
decreases. Notably, the fitting curve based on the $T$-independent
gap bottleneck model fairly matches the experimental data (solid line
in Fig. \ref{fig_Tanl}(b)). Here, the value of $\Delta_{{\rm PG}}=9.8\pm1.6$
meV is obtained.

On the other hand, when looking at the relaxation time, the VOD10
sample data (Fig.~\ref{fig_Tanl}(f)) show a significant difference
from those of the OPD34 sample (Fig. \ref{fig_Tanl}(e)). In the OPD34
sample the relaxation time, $\tau$, remains below $\sim1$ ps in
the temperature range of $T=30-140$ K, despite some slowing-down
with decreasing $T$. In contrast, VOD10 displays a pronounced increase
of $\tau$ with decreasing $T$, following a power-law dependence
of the form $\tau\propto T^{-z}$. 
From the weighted least-squares fit over the temperature range 8--80 K, we obtain an effective exponent of  $z=0.98\pm0.06$ (solid line in Fig. \ref{fig_Tanl}(f)).

In Figs.~\ref{fig_Tanl}(c) and (d), we present the transient reflectivity
analyses for VOD7 and VOD0. The amplitude shows only a weak dependence
on temperature in both samples. However, the relaxation time $\tau$
of the VOD7 sample exhibits a power-law increase (Fig.~\ref{fig_Tanl}(g)),
resembling the behavior observed in the VOD10 sample (Fig.~\ref{fig_Tanl}(f)).
The magnitude of $\tau$ is quantitatively comparable between the
two samples, while a power-law fit (solid line in Fig. \ref{fig_Tanl}(g))
yields a smaller exponent of $z=0.58\pm0.05$.

Finally, the analysis of the VOD0 data reveals an almost $T$-independent
amplitude (Fig. \ref{fig_Tanl}(d)) and relaxation time (Fig. \ref{fig_Tanl}(h)).


\section{Discussion}

The phonon-bottleneck model \cite{kabanov1999} is a two-temperature
model based on an assumption that the thermalization between the high-energy
quasiparticles (HEQP) with energies, $\left|\epsilon-\mu\right|>\Delta$
and high-energy phonons (HEP) with energies $\hbar\omega>2\Delta$
is the fastest relaxation process. Here $\epsilon$ is the quasiparticle
energy, $\mu$ the chemical potential and $\Delta$ the characteristic
energy scale. The factor of 2 between the phonon and quasiparticle
energy scales is assumed to be due to kinematic constraints that favor
quasiparticle recombination with emission of a high energy phonon
and vice versa. As a result the HEQP and HEP transiently attain slightly
elevated temperature with respect to the low-energy degrees of freedom
leading to a characteristic $T$ dependence of the nonequilibrium
HEQP density, $n_{\mathrm{HEQP}}$. By assuming that $\Delta R/R\propto n_{\mathrm{HEQP}}$
and that $\Delta_{\mathrm{PG}}$ is $T$-independent, the model was
successfully used to describe the $T$-dependent transient reflectivity
in the pseudogap state in a wide variety of hole-doped cuprate superconductors
in the underdoped to moderately overdoped phase diagram region \cite{demsar1999superconducting,kusar2005asystematic,toda2011quasiparticle,demsar2001quasiparticle}
including optimally doped Bi-2201 \cite{toda2021ultrafast}. The assumption
of a $T$-independent $\Delta_{\mathrm{PG}}$ is consistent with the
absence of singularities in the heat capacity at the PG crossover
$T$ \cite{lorama1998specific} and $T$-dependent ARPES spectral
intensity \cite{Kondo2011NatPhys,Nakayama2010PhysC} where gap-filling
is observed with increasing $T$.

Turning to the present results, the comparison of the OPD34 and VOD10
data in Fig.~\ref{fig_T} suggests that the negative transient reflectivity
in the VOD10 sample below $T\sim100$ K also originates from the PG
bottleneck response. Indeed, the $T$ dependence of the transient
reflectivity amplitude in VOD10 can be well described by the phonon-bottleneck
model with a $T$-independent gap (Fig. \ref{fig_Tanl}(b)) \cite{kabanov1999},
and its fluence dependence exhibits a saturation behavior {[}Supplemental
Material \cite{SM}, Fig.~S3(b){]}, which can be attributed to the
photoinduced suppression of the PG state \cite{madan2015evidence}.

\begin{figure}[htbp]
\centering \includegraphics[width=0.7\textwidth]{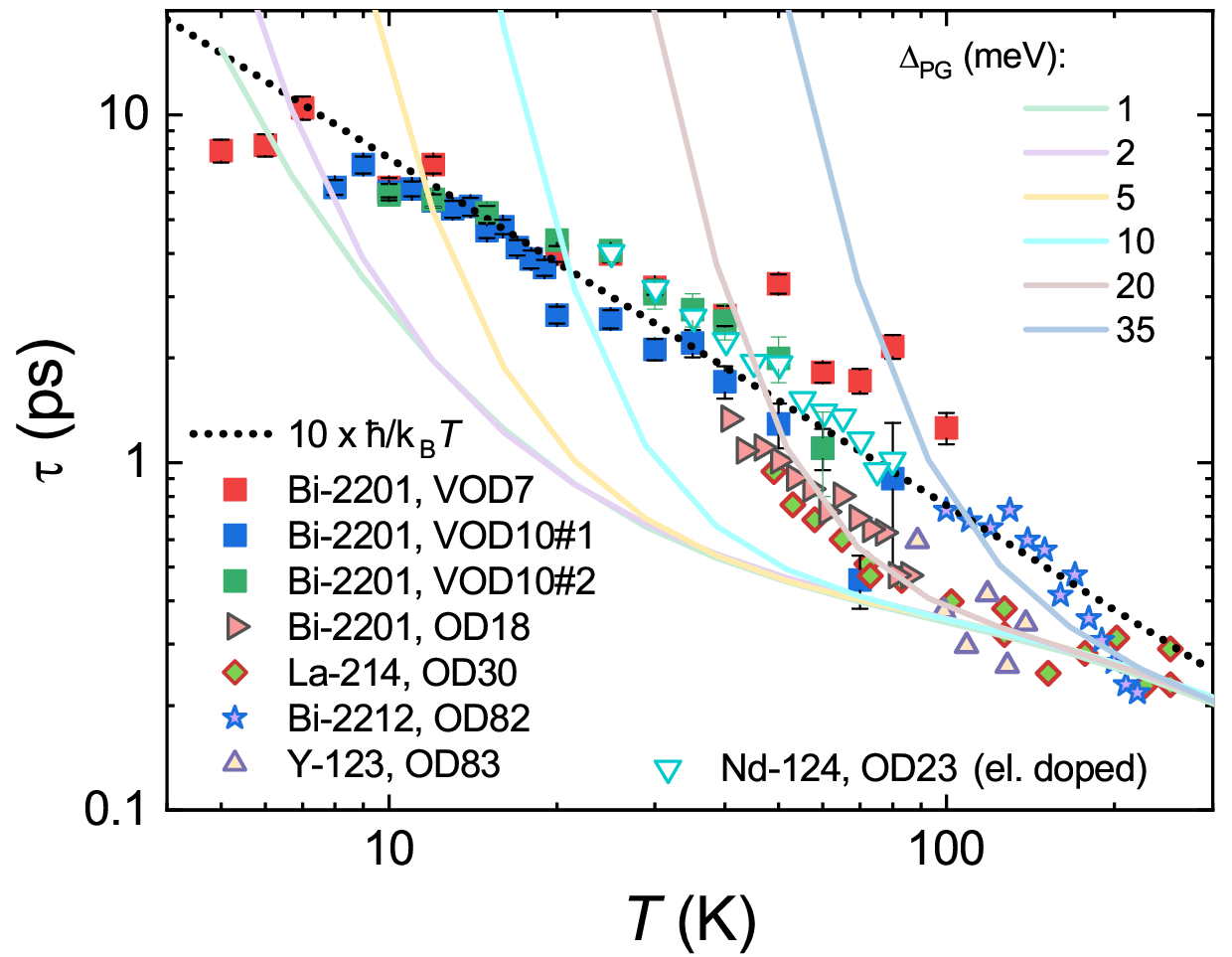}
\caption{Temperature dependence of the normal-state transient-reflectivity relaxation time in various overdoped (OD$x$) cuprates, where $x$ corresponds to $T_{\mathrm{c}}$ in K. For Bi-2201 VOD10, additional data obtained from a different sample piece are also shown to demonstrate reproducibility. The dotted line is a scaled Planckian timescale expected in the quantum critical region. The full lines correspond to the anharmonic bottleneck relaxation time (see Sec.~F of the Supplemental Material~\cite{SM}) scaled to the experimental timescales. The data for La-214, Bi-2212, Y-123, Nd-124 and Bi-2201-OD18 were taken from \cite{kusar2005asystematic}, \cite{toda2014rotational}, \cite{demsar1999superconducting}, \cite{demsar2001quasiparticle}, \cite{hinton2013timeresolved} and \cite{schneider2003timeresolved}, respectively.}
\label{fig:tau-vs-T-all} 
\end{figure}

On the other hand, the relaxation time in the VOD10 sample increases and appears to diverge as $\tau\sim1/T$ (Fig.~\ref{fig_Tanl}(f) and \ref{fig:tau-vs-T-all}). 
The weighted fit yields an effective exponent of $z=0.98\pm0.06$, close to unity, and remains robust for different fitting temperature ranges (see Supplemental Material~\cite{SM}).
A diverging relaxation time, $\tau\sim1/T^{z}$, with $z\sim1.2$, was previously reported in the normal state of electron-doped Nd$_{2-x}$Ce$_{x}$CuO$_{4+\delta}$ (Nd-124) single crystal \cite{hinton2013timeresolved} at $x=0.156$ doping (see Fig. \ref{fig:tau-vs-T-all}). Moreover, a temporal scaling of the transient reflectivity when comparing different $T$ was also observed so the normal-state transient reflectivity was attributed to the dynamics of an order parameter showing critical slowing-down with decreasing $T$. Later, a doping dependent transient reflectivity study \cite{vishik2017ultrafast} in La$_{2-x}$Ce$_{x}$CuO$_{4+\delta}$ thin films showed that at lower dopings the power-law relaxation time $T$ dependence is present only at higher $T$, where the antiferromagnetic correlations are absent. The temporal scaling was not found at lower dopings and the power-law exponent showed an increase with decreasing doping to $z\sim2$ at $x=0.08$.

The normal-state $\tau$ in the hole-doped cuprates has not been analysed
in such detail, in particular in the strongly overdoped region. In
Fig. \ref{fig:tau-vs-T-all} we compile some of the results from literature
focusing on the overdoped phase diagram region. The $\tau\sim1/T$
behavior appears to be present in a wide variety of weakly overdoped
cuprates, albeit in limited temperature ranges. In OD18 Bi-2201 a
positive transient component due to the SC fluctuations emerges below
$T\lesssim40$~K \cite{schneider2003timeresolved} preventing unambiguous
definition of the relaxation time, similar to the present OPD34 sample
(see Fig.~S3a). The $\tau\sim1/T$ scaling is therefore observed
between $\sim40$~K and $\sim85$~K. In OD30 La-214 the SC fluctuations
appear below $\sim50$~K \cite{kusar2005asystematic}. The data between
$\sim50$~K and $\sim150$~K appear consistent with the $\tau\sim1/T$
dependence. In the case of Bi-2212 and Y-123 the SC fluctuations appear
weaker so the low-$T$ limit of the data is $\sim10$~K above the
respective $T_{\mathrm{c}}$s. The $\tau\sim1/T$ trend extends up
to $T\sim170$~K in Bi-2212 and up to $T\sim140$~K in Y-123. No
systematic $\tau\sim1/T$ scaling is found in the optimally doped
samples (see Supplemental Material, Fig.~S9).

In the case of a fully gapped phonon-bottleneck model, the nonequilibrium
HEQP and HEP populations can relax only by anharmonic high-energy
phonon decay or by energy transport into the neighboring unexcited
volume. In the case of the PG, however, the inelastic scattering channels
between HEQPs and low-energy quasiparticles/phonons cannot be excluded
a priori, in particular if they are faster than the anharmonic phonon
decay. To check whether this is the case, we plot in Fig. \ref{fig:tau-vs-T-all}
also the theoretical prediction for the $T$ dependence of the energy
relaxation time for anharmonic high-energy phonon decay taking into
account a realistic Bi-2201 phonon density of states \cite{parshin1996copperatom}.
As expected, the energy relaxation time shows an exponential divergence
at low-$T$, where in the bottleneck regime most of the deposited
energy remains in the nonequilibrium quasiparticle population and
the anharmonic phonon relaxation channel becomes ineffective. At higher
$T$, $k_{\mathrm{B}}T\gtrsim\Delta_{\mathrm{PG}}$, where a significant
amount of the deposited energy is transferred into the high-energy
phonon subsystem, the $T$ dependence flattens out, becoming similar
to the power-law $z\sim0.5$ behavior.

Taking into account the $\Delta_{\mathrm{PG}}$ values obtained from
the amplitude fits, we find that the $T$ dependence of the relaxation
time in the optimally hole-doped cuprates is qualitatively consistent
with the anharmonic relaxation mechanism (see Supplemental Material,
Fig. S9). In the case of weakly hole-overdoped cuprates (see Fig.
\ref{fig:tau-vs-T-all}), however, only OD30 La-214 ($\Delta_{\mathrm{PG}}\sim15$~meV
\cite{kusar2005asystematic}) and possibly Y-123 ( $\Delta_{\mathrm{PG}}\sim20$~meV
\cite{demsar1999superconducting}) data are qualitatively consistent
with the anharmonic relaxation mechanism.

Looking at the strongly overdoped Bi-2201, the anharmonic relaxation
time $T$ dependence is clearly inconsistent with the data in the
VOD10 sample, suggesting that either (i) another HEQP energy relaxation
channel exists with the relaxation rate proportional to $T$, $1/\tau\propto T$,
or (ii) the transient reflectivity (in the strongly overdoped region)
does not correspond to the nonequilibrium quasiparticle density, but
to another (presumably collective) degree of freedom, as proposed
in \cite{hinton2013timeresolved} in the case of Nd-124.

As discussed above, possibility (i) is plausible as the inelastic
HEQP scattering to the states within the PG is less phase-space restricted
than in the case of a $d$-wave superconductor \cite{gedik2004singlequasiparticle}.
Since the density of low-energy states within the PG is depleted with
decreasing $T$ \cite{Kondo2011NatPhys,Nakayama2010PhysC}, an increase
of $\tau$ with decreasing $T$ is expected, consistent with the data.
It is surprising, however, that in the VOD10 sample the depletion
leads to a simple $1/\tau\sim T$ power-law behavior as the in-gap
density of states does not show a $T$-linear behavior in the overdoped
Bi-2201 region \cite{Kondo2011NatPhys}.

A possible origin of the $1/\tau\sim T$ behavior would be proximity
to a QCP at the PGED where a quantum critical fan-shaped region with
the apex at the QCP doping can exist \cite{vojta2003}. This region
is characterized by the coexistence of quantum and thermal critical
fluctuations where the characteristic energy scale is given by $k_{\mathrm{B}}T$.
Inelastic scattering of a photoexcited quasiparticle on such excitations
is connected to the dynamical charge susceptibility \cite{nozi`eres2018theoryof},
$\chi_{\mathrm{c}}(\mathbf{k},\omega,T)$, which is predicted to display
critical scaling properties \cite{vojta2003}. The critical scaling
leads to power-law $T$ dependences of the static uniform susceptibilities.
In the case of a dynamical susceptibility, however, the critical scaling
is more general, $\chi(\mathbf{k},\omega,T)=T^{p}O_{\chi}(\mathbf{k}T^{q},\omega/T)$,
where the scaling function $O_{\chi}(\mathbf{x},y)$ depends on the
$T$-scaled wavevector, $\mathbf{k}$, and $T$-scaled energy, $\omega$.
The wavevector is taken relative to the ordering wavevector while
$p$ and $q$ are some universal exponents. Although the nonequilibrium
energy/population relaxation rate involves some integrals over the
Brillouin zone and energy a power-law $T$ dependence could be expected
only for certain cases depending on the underlying microscopic model.
Since in the case of the putative QCP in the cuprates the microscopic
model and the dynamical charge susceptibility scaling function are
not known a clear connection cannot be deduced. However, the $1/\tau\sim T$
energy relaxation behavior due to HEQP scattering on the quantum critical
fluctuations cannot be excluded despite the lack of generality.

In the context of (i), an inhomogeneous scenario similar to the one
proposed in Ref. \cite{mihailovic2005optical} should also be considered
due to the ubiquitous electronic inhomogeneity in the overdoped Bi-2201
\cite{he2014fermisurface,zheng2017thestudy,Tromp2023NatMat}. In such
a scenario, coexistence of gapped and ungapped regions is assumed
and the excess energy relaxation in the gapped region is attributed
to the energy transport to the ungapped regions. In \cite{mihailovic2005optical},
it was proposed that the energy transport is due to the ballistic
acoustic phonons; however, as discussed above, the anharmonic relaxation
channel slows down exponentially at low-$T$ so the $\tau\sim1/T$
behavior cannot be reproduced.

Alternatively, the energy could be carried away only by HEQPs as the
transport by HEPs can be ruled out due to the exponential suppression
of their density at low $T$. In the case of ballistic transport,
the characteristic relaxation time would be proportional to the characteristic
inhomogeneity length scale, $l_{\mathrm{inh}}$, implying a critical
inhomogeneity-scale behavior, $l_{\mathrm{inh}}\propto1/T$. Taking
the low-$T$ value of $l_{\mathrm{inh}}$ to be a few nm \cite{he2014fermisurface,zheng2017thestudy}
the ballistic velocity of a few hundred m/s is estimated. This would
be unphysically slow, as it corresponds to the thermal velocity of
a quasiparticle with a mass enhancement on the order of 1000 at $T=10$~K.

In the case of diffusive transport, the characteristic relaxation
time is set by the diffusion time, $\tau_{\mathrm{D}}=l_{\mathrm{inh}}^{2}/D$,
where $D$ corresponds to the HEQP diffusion constant. As $D$ usually
increases with decreasing $T$, the $\tau\sim1/T$ behavior implies
an increasing $l_{\mathrm{inh}}\sim\sqrt{D/T}$ with decreasing $T$,
so we can rule out a static, $T$-independent (chemical) sample inhomogeneity
with a constant $l_{\mathrm{inh}}$. Similarly to the previous case,
the estimated value of $D\sim0.003$~cm$^{2}$/s at $T=100$~K is
incompatible with coherent quasiparticle transport~\footnote{Taking the unit cell size, $\sim0.3$~nm, as the lowest possible
mean free path, a velocity on the order of 1~km/s is estimated at
$T=100$~K, corresponding to the thermal velocity of a quasiparticle
with a mass enhancement of a few thousand.}. Diffusive transport would therefore be possible only in the case
of localized HEQP. Moreover, a critical behavior, $l_{\mathrm{inh}}\propto1/T^{\nu}$,
would imply also critical behavior of $D\sim1/T^{2\nu-1}$, with $D\sim1/T$
when $\nu=1$.

We conclude that possibility (i) can consistently reproduce the observed
transient-reflectivity amplitude and relaxation time behavior. While
it is not possible to exactly pinpoint the relevant microscopic relaxation
mechanism, the $1/\tau\sim T$ behavior for the most relevant possibilities
could be naturally connected to the quantum criticality.

Assuming possibility (ii), the transient reflectivity dynamics could
be interpreted in terms of the dynamics of an order parameter that
couples to the optical dielectric function. The $1/\tau\sim T$ behavior
would correspond to the critical slowing down due to a low-$T$ second
order phase transition or possibly a $T=0$ quantum phase transition
at the PGED. As $\tau$ does not diverge at $T_{\mathrm{c}}$ and
the SC response is undetectable even below $T_{\mathrm{c}}$, we can
rule out the SC order parameter.

Moreover, the observed timescale corresponds to, 
\begin{equation}
\tau\sim10\,\tau_{\mathrm{P}},\label{eq:planck}
\end{equation}
indicating a possible connection to the Planckian timescale, $\tau_{\mathrm{P}}=\hbar/k_{\mathrm{B}}T$,
which is predicted to govern the intrinsic dynamics in the quantum
critical region of the phase diagram \cite{sachdev2011quantum} and
is often discussed in the context of the $T$-linear resistivity in
the cuprates \cite{legros2019universal}. While the numerical prefactor
that depends on the details of the underlying model is expected in
\eqref{eq:planck}, the observed value appears somewhat large.

In contrast to the case of Nd-124~\cite{hinton2013timeresolved},
temporal scaling of the transient reflectivity is absent for the present
data. This, however, does not rule out the order parameter scenario
as the scaling, $\Delta R(t)\propto te^{-t/\tau}$ \cite{hinton2013timeresolved},
is expected only in the case of near-critically damped order parameter
dynamics, while in the case of strongly overdamped order parameter
dynamics a two-exponential form, $\Delta R(t)\propto e^{-t/\tau}-e^{-t/\tau_{\mathrm{r}}}$,
which does not possess any intrinsic scaling, is expected.

As shown in Sec.~E of Supplemental Material~\cite{SM}, the absence
of any detectable probe-polarization dependence in our transient reflectivity
signal indicates that the measured response is dominated by an isotropic
($A_{1g}$-like) component~\cite{toda2014rotational}. As an order
parameter must break some symmetry the observed $A_{1g}$-symmetry
response cannot result from a linear coupling of an order parameter
to the probe optical electric field in the form~\cite{stevens2002coherent},
$V\propto\psi R_{ij}E_{i}E_{j}$, where $V$ is the free energy coupling
term, $E_{i}$ an optical electric field component, $\psi$ the order
parameter and $R_{ij}$ a component of a (Raman) tensor that is consistent
with the crystal point group symmetry. The lowest order free-energy
coupling term that can produce an $A_{1g}$-symmetry response must
therefore contain a quadratic form, $\left|\psi\right|^{2}$.

The absence of the linear coupling term is natural for finite-wavevector
order parameters~\cite{schaefer2014collective} as the photons carry
a negligible momentum, $q\approx0$. In contrast, the presence of the
linear term and, as a result, B$_{1g}$- and B$_{2g}$-symmetry responses
is not forbidden for the case of a $q=0$ nematic order parameter.
Despite this, the absence of the B$_{1g}$ and B$_{2g}$ components
does not rule out the nematic order parameter \cite{Fujita2014Science,toda2014rotational,sato2017thermodynamic,Ishida2020JPSJ}
as the B-symmetry responses could be averaged out due to random orientations
of the photoexcited nematic fluctuations.

Considering the coupling to the pump optical pulse the quadratic coupling
can contribute to coherent excitation of the order parameter only
if $\left|\psi(t_{\mathrm{P}})\right|^{2}>0$, where $t_{\mathrm{P}}$
corresponds to the pump-pulse arrival time. In the quantum critical
region, $\psi(t_{\mathrm{P}})$ could be nonzero not only due to the
fluctuations but also due to the presence of a static conjugate field.
Despite the random nature of the fluctuations, the response does
not average to zero due to the quadratic coupling to the probe field. In both cases the magnitude of the response should diminish with increasing $T$
as $\left|\psi(t_{\mathrm{P}})\right|^{2}$ should decrease away from
the critical point.

While the fluctuations scenario would apply for any order parameter
the conjugated field scenario is the most relevant in the case of
an electronic nematic order parameter, where the conjugated field
could be associated with the intrinsic lattice orthorhombicity and/or
an extrinsic strain.

Comparing the quasiparticle relaxation scenario (i) to the order parameter
relaxation scenario (ii), both appear qualitatively consistent with
the data in the VOD10 sample. However, a quantitative theoretical description
of the $T$-dependent $\Delta R/R$ magnitude is lacking for (ii)
while the bottleneck model provides a systematic description of the
data in different cuprate families in the broad range of doping. Unfortunately,
as the probe and pump photon energies lie in the eV range and the
origin of the technique's sensitivity to the low-energy excitations
in the cuprates is not completely understood, it is not possible to
rule out either of the possibilities (i) and (ii). A similar low-$T$ power-law relaxation time behavior is observed
also in the VOD7 sample, but with a different $z\sim0.6$ and different
$\Delta R/R$ sign. Although it is tempting to associate the sign
change with crossing a QCP at the PGED, we note that the sign of the
PG optical response is photon energy dependent in Bi-2212 \cite{Giannetti2011}
with zero crossing close to our probe-photon energy (1.55 eV). In
Bi-2212 the zero-crossing energy shifts with doping resulting in a
similar sign reversal at a photon energy of 1.55~eV in the overdoped
region. The sign change is therefore most likely coincidental, due
to a shift of the zero crossing energy.

Looking at the relaxation time, however, it is plausible to assume
that the low-$T$ relaxation time divergence has the same origin as
in the VOD10 sample since the doping is also in the vicinity of the
PGED. In the VOD7 sample, however, the pump fluence dependence (see
Supplemental Material~\cite{SM}) is quite different from the VOD10
sample, showing only a weak, partial saturation. This, together with
the weak amplitude $T$ dependence, indicates that the contribution
of the metallic response to the transient reflectivity in the VOD7
sample is significant also at low-$T$. The single exponential relaxation
fit procedure therefore does not allow us to reliably extract the
intrinsic relaxation time of the smaller saturable component that
is associated with the PG related dynamics. As the metallic-component
relaxation time is only weakly $T$ dependent the effective relaxation
time obtained from the single exponential fit is expected to show
a weaker $T$ dependence with an apparent $z$ value smaller than
the intrinsic one.

The fluence-dependent slowdown of the relaxation time, observed in
both the VOD10 and VOD7 samples (see Supplemental Material~\cite{SM},
Fig. S3, for details), becomes more pronounced at lower temperatures.
In the VOD7 sample, the slowdown is apparent only for $F\lesssim10~\mu$J/cm$^{2}$,
since at higher fluences the response is dominated by the unsaturated
metallic component. This behavior can be understood as a transient
increase of the effective electronic temperature upon photoexcitation,
which temporarily drives the system away from the QCP.

In the VOD0 sample, the transient reflectivity is predominantly governed
by the metallic carrier dynamics as the saturation is virtually absent
in the full studied fluence range (see Supplemental Material~\cite{SM},
Fig. S3) and the relaxation time is only weakly $T$ and fluence dependent,
consistent with the theoretical analysis of the nonequilibrium-quasiparticle
relaxation in metals \cite{baranov2014theoryof}.

As discussed above, the presence of the $\tau\sim1/T$ behavior in
the hole-overdoped cuprates and in particular in Bi-2201 near the
PGED could be naturally interpreted in the context of the quantum
criticality at the PGED, despite incomplete understanding of the
probe process.
Our results also suggest, contrary to Berben et al.
\cite{berben2022}, that the PGED in Bi-2201 is located at $T_{\mathrm{c}}/T_{\mathrm{c}}^{\mathrm{(max)}}\simeq0.25$,
according to the dashed line in Fig. \ref{fig_pd}. However, to fully
settle this issue, a more detailed transient reflectivity study of
the weakly and moderately overdoped region ($T_{\mathrm{c}}\gtrsim10$~K),
where high quality data are still lacking, would be necessary.

It is also necessary to compare our results with the data from more-direct
low-energy probes. However, as the putative quantum critical point
at the PGED is obscured by the SC dome, a more direct evidence for
its existence is scarce and limited to low-$T_{\mathrm{c}}$ families,
where the SC dome can be suppressed in magnetic field. The most direct
thermodynamic evidence exists in the La-124 family where a peak of
the electronic heat capacity with $T\log(T)$ divergence was observed
\cite{michon2019thermodynamic} near the PGED. Similar behavior was
reported also in Bi-2201 \cite{girod2021normalstate}. In the La-124
system also a strikingly linear low-$T$ resistivity, extending down
to $T\sim1$~K near the PGED, was observed in magnetic field  \cite{Lizaire2021,daou2009lineartemperature}.
In the Bi-2201 system the fully-linear low-$T$ resistance is not
present \cite{berben2022,Lizaire2021} despite the heat capacity $T\log(T)$
divergence \cite{girod2021normalstate}. A fully-linear resistivity
down to $T=20$~K was reported also in overdoped Bi-2212 and down
to a few K in electron doped cuprates \cite{Lizaire2021} with estimated
scattering rates close to the Planckian value. A singular behavior
of the electronic nematic susceptibility near the PGED behavior was
also reported in Bi-2212 \cite{Ishida2020JPSJ}. The interpretation
of the observed $\tau\sim1/T$ behavior in the context of a QCP at
the PGED is therefore consistent with available results from more-direct
low-energy probes.

On the other hand, the linear-in-$T$ component of the zero-magnetic-field
resistivity and the Hall number in Tl-2201 and Bi-2201 \cite{putzke2021}
show a rather smooth evolution across the PGED, suggesting a rather
broad crossover to the Fermi liquid state, well beyond the PGED, even
at $T=0$~K \cite{hussey2018atale}. We note that, unlike in Bi-2212
and La-214, the Lifshitz transition in Bi-2201 occurs near the superconducting
end-point doping rather than at the PGED \cite{Ding2019}, indicating
that the PGED cannot be simply associated with a Fermi-surface topology
change. Considering the ubiquitous presence of doping-induced quenched
disorder and the experimental evidence of the static inhomogeneous
electronic state \cite{he2014fermisurface,zheng2017thestudy,Tromp2023NatMat}
in the overdoped Bi-2201, one should consider smearing of the QCP.
In the case of quenched disorder, theoretical analysis suggests that
an inhomogeneous quantum Griffiths phase (QGP) can exist in a finite
region along the critical control parameter \cite{vojta2013phasesand}.
In the QGP the observables are also expected to display power-law
$T$ dependences, however, the exponents are not universal and depend
on the amount of disorder~\cite{vojta2013phasesand}.

A doping-dependent power-law exponent reported in electron doped La$_{2-x}$Ce$_{x}$CuO$_{4+\delta}$
\cite{vishik2017ultrafast} appears qualitatively consistent with
the QGP scenario. In the present case, the possibly reduced intrinsic
value of $z$ in the VOD7 sample also suggests doping dependence expected
in the QGP. However, the existence of a QGP depends on the details
of the underlying model~\cite{vojta2013phasesand}, which has not
been unveiled for the case of PG, so it is unclear whether the QGP
scenario would be applicable for the cuprates.

\section{Conclusion}

We investigated the temperature-dependent photoinduced transient reflectivity
dynamics in Pb-Bi-2201 single-layer cuprate superconductors, focusing
on the strongly overdoped region of the phase diagram that spans the
pseudogap and superconducting-dome end-point dopings. Near the proposed pseudogap end-point doping, below $T\sim100$~K, we observe
a Planckian-like normal-state relaxation time temperature dependence,
$\tau\sim10\hbar/k_{\mathrm{B}}T$. From the analysis of the literature
data, it appears that such behavior is present also in different hole-overdoped
cuprates, albeit in narrower temperature ranges. We show that such
behavior is incompatible with the anharmonic phonon relaxation described
by the phonon-bottleneck model, which consistently accounts for the
normal-state relaxation dynamics in underdoped and optimally hole-doped
cuprates. Considering different possible relaxation scenarios, we
suggest that the observed behavior is most naturally consistent with a quantum-critical scenario near the proposed pseudogap end-point doping.

Comparing the present behavior to similar results in an electron-overdoped
cuprate \cite{hinton2013timeresolved} suggests universal behavior
with a quantum critical point in the overdoped region; however, more
data in different strongly hole-overdoped cuprates would be needed
to confirm this hypothesis. \\

\section*{Acknowledgments}

The authors declare no conflicts of interest. V.V. K., D. M. and T.
M. acknowledge the financial support of Slovenian Research and Innovation
Agency (research core funding No. P1-0040). Y. T. acknowledges the
financial support of Japan Society for the Promotion of Science (JSPS,
26K08223).

\section*{Data Availability Statement}
The data supporting the findings of this study are available within the article and its Supplemental Material. Additional data supporting the findings of this study are available from the corresponding author upon reasonable request.

\bibliographystyle{unsrt}
\bibliography{YT202503}

\end{document}